# Channel Length Scaling of MoS$_2$ MOSFETs

Han Liu, Adam T. Neal and Peide D. Ye*

School of Electrical and Computer Engineering and Birck Nanotechnology Center,

Purdue University, West Lafayette, IN 47906, United States

*Address correspondence to: yep@purdue.edu

## Abstract

In this article, we investigate electrical transport properties in ultrathin body (UTB) MoS$_2$ two-dimensional (2D) crystals with channel lengths ranging from 2 μm down to 50 nm. We compare the short channel behavior of sets of MOSFETs with various channel thickness, and reveal the superior immunity to short channel effects of MoS$_2$ transistors. We observe no obvious short channel effects on the device with 100 nm channel length (L$_{ch}$) fabricated on a 5 nm thick MoS$_2$ 2D crystal even when using 300 nm thick SiO$_2$ as gate dielectric, and has a current on/off ratio up to ~10$^9$. We also observe the on-current saturation at short channel devices with continuous scaling due to the carrier velocity saturation. Also, we reveal the performance limit of short channel MoS$_2$ transistors is dominated by the large contact resistance from the Schottky barrier between Ni and MoS$_2$ interface, where a fully transparent contact is needed to achieve a high-performance short channel device.

In the past decade, as scaling of silicon based transistor has approached its physical limit, intensive efforts in finding alternative channel materials for future logic devices beyond 10 nm node have been made with the main focus on Ge and III-V materials because of their superior carrier mobility.[1-4] The discovery of graphene has unveiled another material family with layered structures, which includes boron nitride, topological insulators such as $Bi_2Te_3$ and $Bi_2Se_3$, and transition metal dichalcogenides like $MoS_2$, $WS_2$, and $NbSe_2$.[5-12] Though graphene, a fascinating two-dimensional (2D) crystal, has shown a superior carrier mobility of up to 200,000 $cm^2/V \cdot s$, its zero bandgap property limits its application to logic devices as graphene transistors cannot have high on/off ratios.[13] As opposed to the semi-metal graphene, transition metal dichalcogenides (such as $MoS_2$), as another type of layered structure material, have shown great potential in device applications due to their satisfied bandgaps, thermal stability, carrier mobility, and compatibility to silicon CMOS process.[11] In order to realize high performance $MoS_2$ or some other transition metal dichalcogenide MOSFETs, three major issues must be solved: 1) how to deposit a high-quality dielectric on 2D crystal, 2) the fabrication of low-resistivity metal-semiconductor junction to be used as device contacts, and 3) the elimination of short channel effects. Although the high-k dielectric has been successfully demonstrated in several previous reports,[11,14,15] the interface between high-k dielectric still needs to be systematically studied. Also, as the 2D material cannot be effectively implanted due to the nature of ultrathin body, the contact resistance ($R_c$) is mostly determined by the Schottky contact at the $MoS_2$/metal interface. This contact resistance at the MoS$_2$/metal junction is much larger than for contacts of other metal/low-dimensional systems (e.g. graphene or carbon nanotube) due to the enlarged Schottky barrier height (SBH) induced by the much wider bandgap of MoS$_2$. Thus, to find a metal or alloy having the correct work function located near or even into the conduction (valence) band edge for n-type (p-type) transistors, becomes significantly important. It could be very difficult due to the metal induced gap states at the MoS$_2$/metal interface. The third issue is related to dimension scaling and the transistor density of a single chip. For logic applications, the performance limits of MoS$_2$ transistors associated with channel length scaling must also be investigated. Classical discussions on short channel effects are mostly based on silicon MOSFETs. However for the MoS$_2$ transistors, the origins and behaviors of short channel effects could be slightly different from silicon MOSFETs simply because a) the MoS$_2$ transistors are fundamentally majority carrier devices with carrier accumulation for "ON" state while silicon MOSFETs are minority carrier devices with carrier inversion for "ON" state; b) the source/drain areas of MoS$_2$ transistors are not heavily doped, and they are simple metal/semiconductor junctions; and c) the characteristic length for short channel MoS$_2$ transistors is smaller due to the low dielectric constant of MoS$_2$.

**Results and Discussions**

We fabricated sets of MoS$_2$ MOSFETs with various channel length. Each set was fabricated on the same rectangular MoS$_2$ flake, so the scaling effect can be directly observed and compared without needing to correct for geometry and thickness variations. The flakes were mechanically exfoliated from a bulk ingot as described in previous studies and transferred to a heavily doped Si substrate with a 300 nm $SiO_2$ capping layer. The heavily doped silicon substrate serves as the global back gate and the $SiO_2$ as the dielectric.[12] More than 10 sets of devices were fabricated. Due to the variation in geometry, including the flake size and thickness, as well as defects density level between different flakes, it is difficult to compare the device performance directly. Here we select the $MoS_2$ devices fabricated on one ~5 nm thick crystal which corresponds to ~6 layers with a rectangular shape as a representative. We didn't reduce the thickness of the $MoS_2$ crystal to a single layer because the larger bandgap of the monolayer may have reduced electron mobility.[16] The schematic and corresponding optical microscope image of the 5 nm thick devices are shown in Figure 1(a) and 1(b), and have various channel lengths from 2 μm down to 100 nm, as defined by electron beam lithography. Metallization was performed by electron beam evaporation afterwards. The width of the contact bars are 500 nm. To realize high performance short channel devices, one of the major issues is to reduce the source/drain contact resistance. In addition to previously demonstrated Au or Ti/Au contacts,[11] we also used Ni/Au as the source/drain metal. No annealing was performed after lift-off process. The Ni/Au contact resistance was extracted using the two-terminal transfer length method (TLM) measurement of the same structure, as shown in Figure 1(c). We extracted the low-field ($V_{ds}$=50mV) contact resistance from devices with larger channel length (>500 nm), which is much larger than the carrier mean free path in the channel,[20] so that the electron transport can be considered as entirely in the diffusive regime. The measurement was performed at room temperature. The contact resistance shows a strong dependence on the back gate bias, as the $MoS_2$ crystal is electrically-doped under high gate bias, leading to a smaller contact resistance. The smallest $R_c$ measured in the $Ni/MoS_2$ junction is 4.7±0.5 Ω·mm at 50V back gate bias, and increased to 18.0±5.9 Ω·mm at zero back gate bias. The contact resistance is about a factor of 40 larger than the Pd/graphene contact,[17] for the absence of a Schottky barrier at metal/graphene junction. Note that the error bars on the left side are significantly larger than those on the right, where the channel is heavily doped. This is attributed to a larger contact resistance on $MoS_2$ at lower gate bias, leading to a larger absolute error, which is also observed in former graphene TLM study.[17] Generally, the gate dependence of $R_c$ can be attributed to two reasons. One is the existence of a Schottky barrier at the metal/semiconductor interface, as gate bias would change the tunneling efficiency due to band bending at the metal/semiconductor interface. The other is the electrical doping of the semiconductor, as happens with graphene.[17,18] As a comparison, an $Au/MoS_2$ TLM structure is also fabricated on another flake with similar thickness (~6 nm), with its contact resistance shown in the same figure. Despite the variance between the two flakes, our results reveal similar contact resistances under the same gate biases. The Schottky barrier between at the $MoS_2/Au$ junction is also observed in a previous temperature dependence study, where the measured mobility showed strong degradation at low temperatures. This could be understood as an increasing contact resistance due to the reduced thermionic emission.[19]

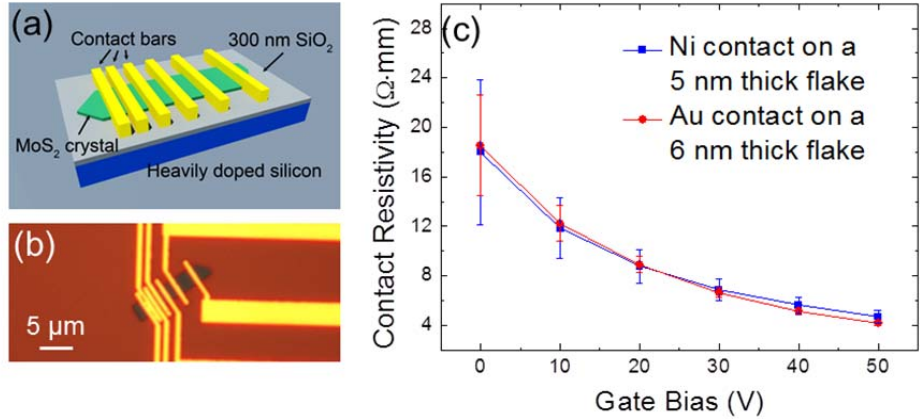

Figure 1: (a) Schematic diagram of the back-gated $MoS_2$ MOSFETs in a TLM structure. Heavily doped silicon is used as the back gate and 300 nm $SiO_2$ as the gate dielectric. Ni/Au or Au is used as source/drain contact. (b) Optical microscope image of one of the fabricated devices. Scale bar is 5 μm. (c) Comparison of contact resistance of Ni/Au on a 5 nm thick device and Au on a 6 nm thick device.

We examined the transistor characteristics of both long-channel and short-channel $MoS_2$ MOSFETs. The study was carried out on the same sets of devices. Considering that the monolayer has been shown to have a larger bandgap and hence a lower mobility and larger contact resistance, we fabricated the devices on a few-layer crystal for a better tradeoff between the on/off ratio and device performance. Note that the dielectric constant of $MoS_2$ is only around 3.3, according to a previous theoretical study,[20] and a 5 nm thick crystal would be thin enough for short channel devices to turn off completely. Figure 2 shows the transfer and output curves for the 2 μm and 100 nm channel length ($L_{ch}$) devices. Drain current saturation is observed in short channel devices as shown in Figure 2(d). Because of their large bandgap of 1.2 eV, these devices, unlike graphene, can be easily turned off. Even though the thickness of gate dielectric is extremely large (300 nm), which results in a much degraded electrostatic control, still no evident short channel effects were observed with channel lengths down to 100 nm. For this short channel device, the on-current is reaching 70 mA/mm at $V_{ds}$=1 V, and the current on/off ratio is over $10^7$ for $V_{ds}$=1 V, and is able to maintain an on/off ratio of $10^9$ at $V_{ds}$=0.1 V. Benefiting from its ultrathin body, the on/off ratio doesn't drop much compared to the 2 μm long device which has a current on/off ratio up to ~$10^{10}$, showing good immunity to short channel effects. Note that significant short channel effects could be observed on other planar devices, such as InGaAs or Ge, when the gate length was scaled down to 150 nm.[21] The intrinsic mobility extracted from the 2 μm long device is ~28 cm$^2$/V·s. It could be further increased up to several hundred by dielectric passivation on the top.[11,14] Our observation of transistor behavior without evident short channel effects with 300 nm SiO$_2$ indicates that the enhancement of electrostatic control by reducing the gate dielectric thickness down to several nanometers would significantly push the scaling of channel length down to sub-10 nm for MoS$_2$ devices. This is beyond the range of conventional semiconductors. The superior immunity to short channel effects of MoS$_2$ not only originates from its ultrathin body nature and junction-less contacts, but is also due to the low dielectric constant of MoS$_2$ itself. The characteristic length of short channel transistors with planar structures is: [22]

$$\lambda = \sqrt{\frac{\varepsilon_s}{\varepsilon_{ox}} t_s t_{ox}}$$

where $\lambda$ is the characteristic length, $\varepsilon_s$ and $\varepsilon_{ox}$ are the permittivity of semiconductor and gate oxide, $t_s$ and $t_{ox}$ are the thickness of semiconductor channel and gate oxide. The characteristic length for this 5 nm thick MoS$_2$ transistor is 35.6 nm, much shorter than the channel length of our shortest device. If the 300 nm SiO$_2$ gate oxide is replaced by 6 nm $HfO_2$ with an equivalent oxide thickness (EOT) of ~ 1 nm, we can expect the characteristic length would be reduced to only 2 nm, which is far beyond the technical consideration of 10 nm node with alternative channel materials for logic applications. This formula was first proposed by Yan *et al* to calculate the characteristic lengths of silicon or other bulk semiconductors, where the carrier transport is almost isotropic.[22] For layered structures, the layer-to-layer transport is more resistive than in-plane transport. Therefore, the effective $t_s$ for 2D crystals could be even smaller, further reducing characteristic length of $MoS_2$ transistors.

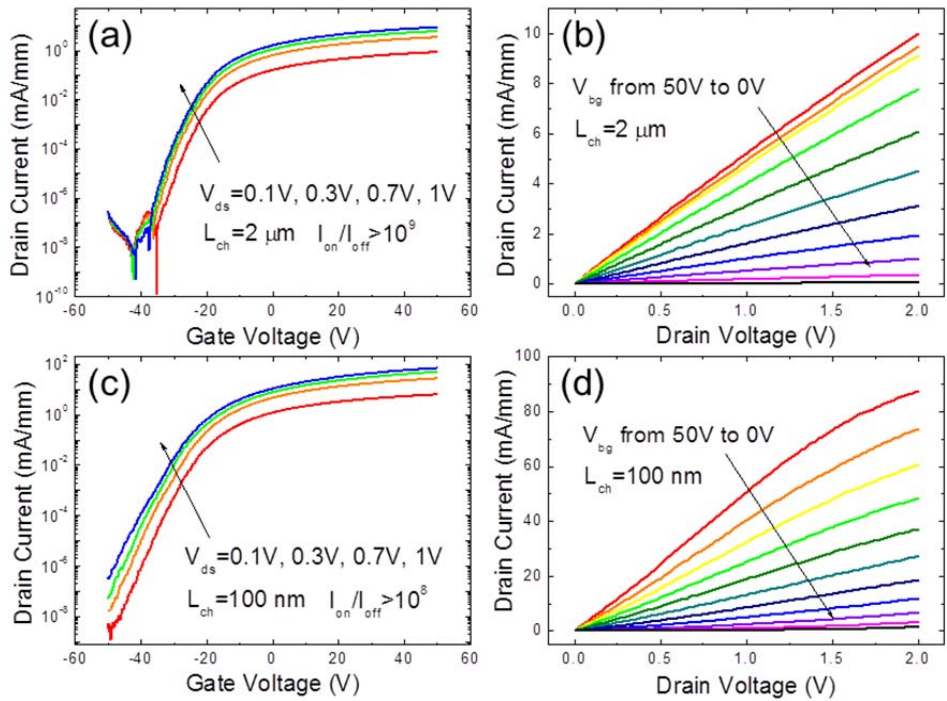

Figure 2: (a) Semi-log plot of the transfer characteristics of a 2 μm long device fabricated on a 5 nm thick $MoS_2$ crystal. Drain voltage is applied from 0.1 V to 1 V with a 0.3 V step. Back gate voltage is swept from -50 V to 50 V. (b) Output characteristics of the same device. Back gate voltage is applied from 50 V to 0V with a -5 V step. Drain voltage is swept from 0V to 2V. (c)(d) Transfer and output characteristics from a device with $L_{ch}$=100 nm.

To make a comparison of the short channel effects related to the MoS$_2$ flake thickness, we fabricated another set of devices on a 12 nm thick MoS$_2$ crystal. The characteristic length of this transistor is calculated to be ~55.2 nm, on the same 300 nm SiO$_2$ as back gate dielectric. We further scaled the channel length down to 50 nm, so that the channel length would be comparable to $\lambda$. The transfer and output characteristics of the device with 50 nm channel length are presented in Figure 3(a) and 3(b). We start to observe obvious short channel effects. The drain current on/off ratio ($I_{on}/I_{off}$) drops down to ~$10^7$ at $V_{ds}$=0.1 V, and ~$5\times10^4$ at $V_{ds}$=1 V. A severe drain induced barrier lowing (DIBL) is also observed. The upward bending in the output characteristics in Figure 3(b) at high drain biases also indicates a degraded electrostatic control from the gate. The channel length dependent $I_{on}/I_{off}$ and DIBL of the two sets of devices are plotted in Figure 3(c) and 3(d). For the set of devices with 5 nm thick MoS$_2$ crystal, the $I_{on}/I_{off}$ ratio is nearly constant, with a minor decrease as the channel becomes shorter, while the total change remains within one order of magnitude of its long-channel value. The set of devices on the 12 nm thick crystal are observed to have a lower $I_{on}/I_{off}$ ratio as expected, following the same slightly decreasing trend with scaling down, until the channel length approaches the characteristic length, where it experiences a sharp drop down to less than $10^5$. Similar behavior is also observed in the DIBL. Due to the weaker electrostatic control from the global back gate with 300 nm gate dielectric, the DIBL is relatively large even at long channel devices compared to top gate devices with sub-10 nm high-k dielectric. We can observe from Figure 3(d) that the DIBL for the sets of devices fabricated on the 5 nm thick flake is smaller than that of devices with the 12 nm thick flake. At long channel lengths, (e.g. $L_{ch}$=2 μm), it is ~4V/V for the thinner devices and ~10V/V for the thicker ones. The DIBL from both sets of devices experience a rapid increase once the channel length approaches the characteristic length as the typical short channel effects.

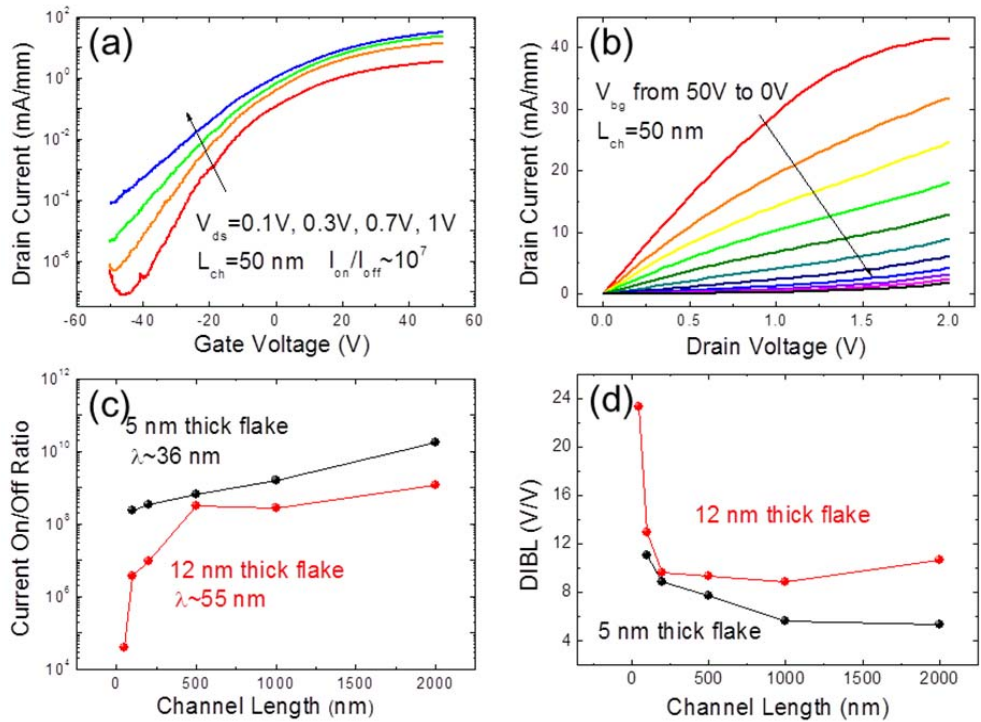

Figure 3: (a)(b) Transfer and output characteristics from the device of 50 nm channel length fabricated on the 12 nm thick $MoS_2$ crystal. (c) Channel length dependent current on/off ratio of the sets of devices fabricated on the 5 nm and 12 nm thick crystals, respectively. The on/off ratio is estimation due to the blurry "OFF" state current. (d) DIBL extracted from the transfer characteristics of sets of devices on 5 nm and 12 nm $MoS_2$ crystals.

For long channel devices, where $L_{ch}$ is much larger than the length of electron mean free path ($L_{mfp}$), the transistors are fully operated in the diffusive regime, where field-effect mobility would remain constant, while the maximum drain current and the transconductance ($g_m$) keep increasing with continuous scaling, which is inversely proportional to $L_{ch}$. We plot both extrinsic/intrinsic field dependent mobility and maximum on-state current at $V_g$=50V and $V_{ds}$=2V for all devices with various $L_{ch}$ in Figure 4(a) and 4(b). We first extracted the peak transconductance by differentiating the transfer curve, and then calculated the extrinsic field-effect mobility by simply using the equation $g_m=\mu_n C_{ox} W/LV_{ds}$, where $\mu_n$ is the electron mobility, $C_{ox}$ is the MOS capacitance, $W$ and $L$ are the width and length of the channel, and $V_{ds}$ is the drain voltage. The intrinsic values of the field-effect mobility are further corrected by calculating the channel resistance at the voltage point where transconductance is at its peak, and then amended the drain voltage $V_{ds}'=V_{ds}(R_{tot}-R_c)/R_{tot}$, where $V_{ds}'$ is the actual drain voltage applied on the channel, $R_{tot}$ is the total resistance and $R_c$ is the contact resistance, as both $R_{tot}$ and $R_c$ are known. Here, we assume diffusive transport for all sets of devices regardless of their channel lengths, and thus we can observe the change of field-effect mobility at different channel length scale. We see from Figure 4(a) that in the long channel regions ($L_{ch}$>500 nm), $\mu_n$ remains constant at around 28 cm$^2$/V·s. With further scaling, $\mu_n$ starts to decrease, and drops to around 17 cm$^2$/V·s at 100 nm channel length. Also, we learn from the classical square-law model that the drain current is inversely proportional to channel length, which means, the $I_d$-$L_{ch}^{-1}$ relationship should present a linear characteristic. However, in Figure 4(b), as indicated by the red dashed line, this linear relationship applied only at long channel region ($L_{ch}$>500 nm). With continuous scaling down, it comes to saturate at ~90 mA/mm at $L_{ch}$=100 nm. The decrease of field-effect mobility and non-linear scaling of drain current are attributed to two reasons. One reason is the substantial contact resistance which does not scale with channel length but is present in the device when the contact resistance is comparable to channel resistance. The second reason, if we don't consider the contact resistance, is that mobility decreases since the carriers are approaching their saturation velocity at shorter channel lengths.[23] In general, the electric field in the channel is reversely proportional to the channel length, leading to higher carrier velocities at reduced channel length, as defined by v=μE. As a result, the drain current increases with reduced channel length, while field-effect mobility remains constant. However, at very short channel lengths, the velocity of the carrier is getting saturated even with increase electric field. Therefore, as the velocity is approaching saturation with increased electric field, the field-effect mobility calculated from the same formula would result in a decreased number, as well as the drain current gets saturated at the same time. As we can see in the figure, the field-effect mobility shows a descending trend when $L_{ch}$ is less than 500 nm, indicating that these short channel transistors with $L_{ch}$ < 500 nm are showing carrier velocity saturation behaviors, which is consistent with what was observed in conventional short channel silicon devices.

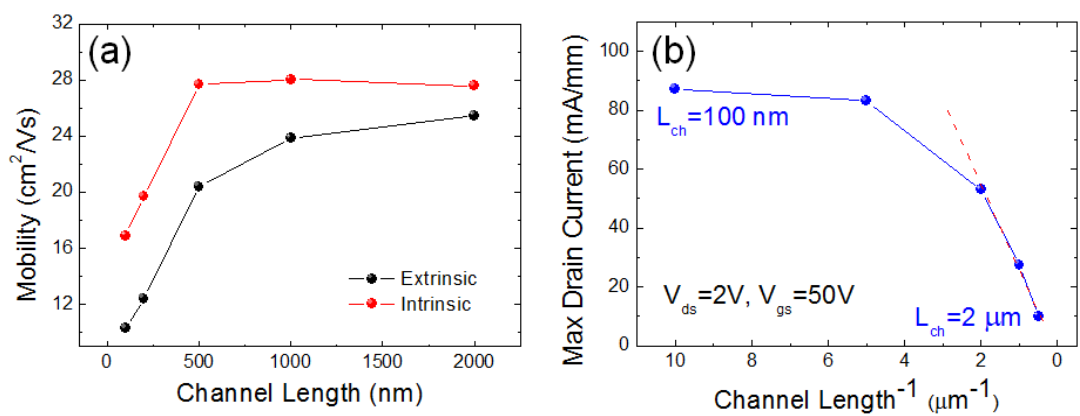

Figure 4: (a) Channel length dependent field-effect mobility from the set of devices fabricated on the 5 nm thick $MoS_2$ crystal using diffusive transport equations. (b) Magnitude of "ON" state drain current measured at 50V back gate bias and 2V drain bias on the set of devices fabricated on the 5 nm thick $MoS_2$ crystal. Minor threshold voltage ($V_T$) shift is neglected here. The linear dependence in diffusive region is indicated by the red dashed line.

Finally, we compare the output curves of all sets of devices with various channel lengths in their "ON" state. The $I_d$-$V_d$ characteristics in Figure 5 show how scaling affects the on-resistance ($R_{on}$) and drain current at a fixed gate voltage. The $R_{on}$ has contributions from both the channel resistance and the contact resistances. Ideally, $R_{on}$ should decrease linearly with the scaling of channel length. Such a decrease would be characterized by the increase of the $I_d$-$V_d$ slope at low drain bias, where acoustic phonon scattering is dominant.[24,25] However, this trend fails to hold for short channel devices, as seen by comparing the 100 nm and 200 nm devices. For these short channel devices, $R_{on}$ saturated at ~20 Ω·mm at $V_{ds}$<0.5 V. The actual saturated $R_{on}$ would be lower than this value if we consider run-to-run variations in measurement of the threshold voltage ($V_T$), mostly originating from fixed charges in the oxide.[14] In the meantime, we can read from Figure 1(c) that the contact resistance at 50 V gate bias is around ~5 Ω·mm, which equals to 10 Ω·mm for the total contact resistance. This indicates that $R_{on}$ with $L_{ch} \lesssim$ 200 nm is largely composed of the contact resistance, since the channel resistance is be small or even negligible at these channel lengths. $R_c$ dominance is not desirable for short channel transistors, which can be understood from two ways. One is that it makes further aggressive scaling worthless, for the $R_{on}$ no longer has a substantial dependence on the channel and thus the drain current doesn't increase with scaling. And the other reason is that the drain voltage applied would be mostly loaded by the contact resistance instead of the channel resistance and result in a degraded lateral electric field in the channel, making it difficult for the drain current to reach the current saturation regime unless a very large drain voltage is applied, as we observe in Figure 5. Slight current saturation occurs only at 100 and 200 nm channel length near the 2V drain bias. That is to say, due to the large contact resistance at metal/$MoS_2$ junction, which overshadows the gradual change in the channel resistance, these transistors are operating way below their intrinsic performance limit. For $MoS_2$, the charge neutrality level is aligned at the vicinity of the conduction band, making it easy to fabricate an n-channel MOSFET on this material.[26]

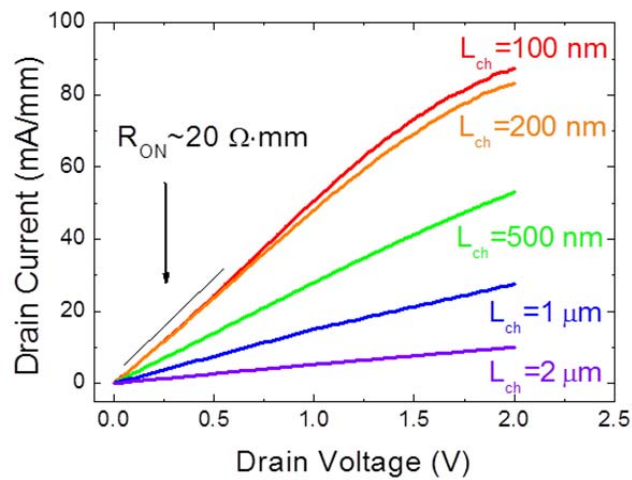

Figure 5: Output curves at the same 50 V gate bias from a series of transistors on the same 5 nm thick flake with different channel length. The 100 nm and 200 nm device exhibit similar slopes, indicating a similar on-resistance at reduced channel lengths. The performance of the transistors at 100 and 200 nm channel length is mostly limited by the dominant contact resistance.

We assume the metal contact on MoS$_2$ is pinned or at least weakly pinned at 1-2 hundred meV below the conduction band edge, so to understand that both high work function metal (Au, Ni) and low work function metal (Ti) work well for a reasonable quasi-ohmic contact to MoS$_2$ at room temperature, as shown in this and some other previous studies.[11,14,19] In order to investigate the metal contact on MoS$_2$, temperature dependent current density of Ni/MoS$_2$ Schottky contact is measured to determine the effective Schottky barrier height (SBH). The current of the pure Schottky diode is $I = A^* A T^2 \exp(-q\Phi_B/kT)\left[\exp(qV/nkT)-1\right]$, where $A^*$ is the Richardson constant, $A$ is the area of the metal contact, $T$ is the temperature, $q$ is the electron charge, $\Phi_B$ is the effective SBH, $k$ is the Boltzmann constant, $n$ is the ideality factor, and $V$ is the applied forward voltage. For Schottky diodes biased at $V>3kT/q$, the effective SBH $\Phi_B$ can be accurately extracted from the slope of temperature dependence of $\ln(I/T^2)$ versus $T$ as shown in the inset of Figure 6. Fig. 6 plots the extracted SBHs as a function of back gate bias, which electro-statically dopes MoS$_2$. It is not surprised to see that the effective SBH is sensitive to the doping concentration of MoS$_2$ and decreases from maximum 100 meV to near 0 eV (Ohmic). The existence of a small Schottky barrier between metal and MoS$_2$ is confirmed by this experiment. A transparent Ohmic contact scheme on MoS$_2$ must be developed before we can realize high-performance short channel MoS$_2$ transistors with a competitive contact resistance with the state-of-the-art semiconductor device technology, i.e., 0.1 Ω·mm. A comprehensive study of the metal/MoS$_2$ interface is critical to understand the origins and limitations for the metal/MoS$_2$ contacts.

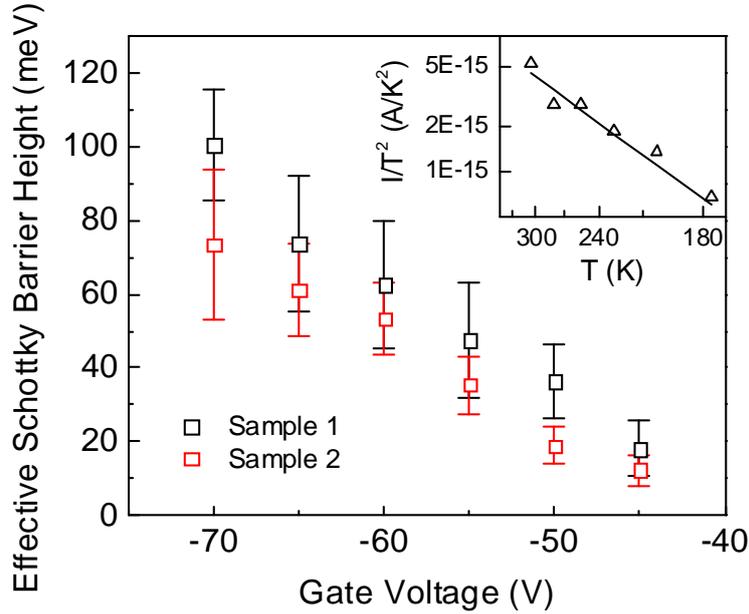

Figure 6: Effective Schottky barrier height of 30nm/30nm Ni/Au contact to MoS$_2$ as a function of back gate voltage extracted via the activation energy method on two samples. Error bars are determined from the standard errors of the linear fitting used to extract the barrier height. Inset: Richardson Plot of $I/T^2$ versus $T$ at back gate bias of -70V. Solid line is the linear fit used to extract the Schottky barrier height.

**Conclusion**

We have studied the device performance of MoS$_2$ short channel transistors. Despite our devices being fabricated on a 300 nm thick SiO$_2$ gate dielectric, the superior immunity to short channel effects down to 100 nm channel length has been demonstrated. We observe a severe decrease of current on/off ratio as well as a sharp increase in DIBL for the device with channel length smaller than the characteristic length. By calculating the characteristic length, we have revealed that the channel length can be aggressively reduced to sub-10 nm if we substitute the currently used

300 nm SiO$_2$ with a state-of-art high-k dielectric. Transport studies are also performed, where the field-effective mobility decrease and maximum current saturation are observed at short channel lengths attributed to carrier velocity saturation. From on-resistance saturation with the channel scaling, we revealed that the large contact resistance, due to the existing Schottky barrier at metal/MoS$_2$ contacts, impedes the short channel device performance. The effective Schottky barrier heights of ~100 meV or less between Ni and MoS$_2$ interface are experimentally determined. In order to realize high-performance MoS$_2$ short channel transistors, a comprehensive study on metal/MoS$_2$ junctions is urgently needed and a transparent Ohmic contact scheme on MoS$_2$ and other 2D crystals [27,28] needs to be developed.

**Method**

The MoS$_2$ flakes were mechanically exfoliated with 3M scotch tapes from a bulk ingot purchased from SPI Inc. A heavily p-doped silicon wafer (0.01-0.02 Ω·cm) with 300 nm SiO$_2$ capping layer was used as back gate and gate dielectric. After flake transfer, the samples were soaked in acetone for overnight to remove the tape residues on SiO$_2$ substrate, followed by methanol and isopropanol rinse. The thickness of the flakes was measured by Dimension 3100 AFM systems. TLM and MOSFETs structures were defined by electron beam lithography, followed by the electron beam evaporation of Ni/Au for 30/50 nm or only 50 nm Au for different sets of devices with the deposition rate of ~1Å/s. After metal deposition, these samples were transferred to Remover PG solution for lift-off process. Electrical characterizations were carried out with Keithley 4200 system at room temperature.

**Acknowledgement**

The authors would like to thank Zhihong Chen, Jiangjiang Gu, Min Xu and Nathan J. Conrad for valuable discussions.

**Table of Contents (TOC) Graphic**

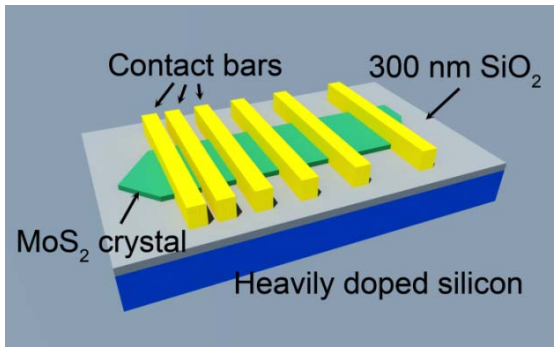 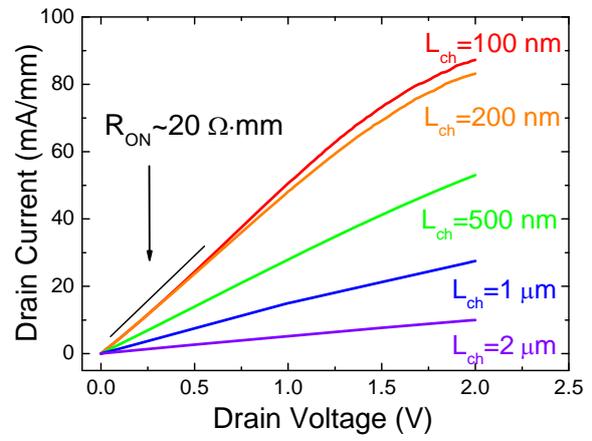